# Magnetic Field Melting of the Charge-Ordered State of $La_{0.5}Ca_{0.5}MnO_3$: A Local Structure Perspective

T. A. Tyson[1], M. Deleon[1], M. Croft[2], V. G. Harris[3], C.-C. Kao[4], J. Kirkland[3] and S.-W. Cheong[2]

[1]Department of Physics, New Jersey Institute of Technology, Newark, NJ  07102
[2]Department of Physics and Astronomy, Rutgers University, Piscataway, NJ 08854
[3]Materials Physics Branch, Naval Research Laboratory, Washington DC 20375
[4]Brookhaven National Laboratory, Upton, Long Island, NY 11973

## Abstract

The local structure about the Mn site in the half doped system $La_{0.5}Ca_{0.5}MnO_3$ was measured in magnetic fields up 10 T to probe the melting of the charge ordered state. Examination of the Mn-O and Mn-Mn correlations reveal three distinct regions in the structure-field diagram.  A broad region with weak field dependence (mainly antiferromatnetic phase below 7.5 T), a narrow-mixed phase region near ~ 8.5 T followed by a ferromagnetic phase region with strong field-structure coupling.  At high field the Mn-O radial distribution becomes Gaussian and the Mn-Mn correlations are enhanced - consistent with the dominance of a ferromagnetic phase.  The exponential change in resistivity in the first region (observed in transport measurements) is dominated by the reordering of the moments on the Mn sites from CE type antiferromagnetic to ferromagnetic order with only a weak change in the local distortions of the $MnO_6$ octahedra.

**PACS:** 78.70.Dm, 61.10 Ht, 75.47.Lx, 87.64.Gb

# I. Introduction

The CMR system ($Re_{1-x}A_xMnO_3$, where A is a divalent cation (such as Ca, Sr or Ba) and Re is a trivalent rare earth metal) has intriguing properties manifested by complex phase diagrams which an be tuned by doping, internal and external pressure and electric and magnetic fields [1, 2, 3]. As a function of these parameters and x (doping level), the system can be forced between metallic and insulating regions. For the classic $La_{1-x}Ca_xMnO_3$ system, the reference state (x=0) is an antiferromagnetic insulator. Ca doping system produces a ferromagnetic conducting system for x~0.2 to x~0.5 and for x > 0.5 and antiferromagnetic insulating phases are found. These materials are rich in intriguing phenomena and are an excellent test bed for studies of systems in which electron-electron and electron phonon interactions can not be neglected [4].

The half-doped system systems $Re_{0.5}A_{0.5}MnO_3$ are at the boundary of the ferromagnetic and charge ordered (CO) insulating region of the phase diagram. These systems are characterized by the onset of ferromagnetic ordering (dominated by the double exchange) followed by a significant loss of bulk magnetization at lower temperature where super exchange dominates. The balance between superexchange and double exchange can be shifted by application magnetic field [5, 6, 7, 8, 9, 10, 11] and Mn site doping [12]. In $La_{0.5}Ca_{0.5}MnO_3$, neutron diffraction measurements found evidence for distinct structural phases corresponding to ferromagnetic and antiferromagneric regions co-existing at low temperature [13, 14]. Cooling in high magnetic field is found to stabilize the ferromagnetic phase and reverse the volume ratios of the two phases [13,15].

The coupling to magnetic field suggest very closely lying energy states of different magnetic structures. Indeed, in $La_{0.5}Ca_{0.5}MnO_3$, the coexistence of mixed phases of insulating and metallic region has been found to be a common feature of manganites and other strongly correlated systems due to the small atomic changes requires to move from one phase to another [16, 17]. Intertwined ferromagnetic (with full saturation magnetic moments per Mn site) and antferromagnetic region have been observed on a nanoscale [17].

In order to explore the nature of the field induced melting of the charge ordered state in $La_{0.5}Ca_{0.5}MnO_3$ on the application of magnetic field a local structural perspective is needed. X-ray absorption spectroscopy enables one to measure the average structure about a specific atom unbiased by the need to impose a long-range structure. This approach enablen studies of the local distortions in the $MnO_6$ octahedra near x ~ 0.3 in $La_{1-x}Ca_xMnO_3$ which revealed the loss of the Jahn Teller distortions on going into the low temperature metallic phase [18].

The $La_{0.5}Ca_{0.5}MnO_3$ system is characterized by the onset of ferromagnetism near 225 K ($T_N$) followed by a near collapse of the net magnetization below $T_N$ ~ 155K. The low temperature phase is has been characterized as a charge ($Mn^{3+}$ an $Mn^{4+}$) and orbital ordered state. In order to understand the structural details of the magnetic field melting of the charge ordered state, we have performed magnetic field dependent x-ray absorption measurements. We have found that the Debye-Waller (DW) factor for the Mn-O distribution is reduced by 40 % between 0 and 10T. Examination of the Mn-O and Mn-Mn correlation reveals three distinct regions in the structure-field diagram. A broad region with weak field dependence (mainly antiferromatnetic phase below 7.5 T), a narrow-mixed phase region near ~ 8.5 T followed by a ferromagnetic phase region with strong field-structure coupling. The Mn-O radial distribution becomes Gaussian and the Mn-Mn correlations are enhanced in the high field region- consistent with the dominance of a ferromagnetic phase. The exponential change in resistivity observed in the first region is found to be dominated by the reordering of the moments on the Mn sites from CE type antiferromagnetic to ferromagnetic order. The more rapid reduction in resistivity at high field is seen to be the result the delocalization of JT trapped electrons as the magnetic field destabilizes the local distortions.

## II. Experimental Methods

The polycrystalline samples of $La_{0.5}Ca_{0.5}MnO_3$ and $CaMnO_3$ were prepared by solid state reaction in air as described in Ref. [19] and extensive studies of the transport, magnetic and zero field structural studies were presented in Ref. [20]. The Néel and Curie temperatures in for the samples were 225 K and 155 K, respectively. X- ray absorption samples (XAFS) were prepared by grinding the sample to form a 400 mesh powder and brushing it onto scotch tape. Layers of tape were stacked to produce a uniform sample for transmission measurements with jump μt ~ 1. A custom designed Oxford Instruments 10T split-coil horizontal field superconducting magnet was used for the measurements [21]. All measurements reported were performed with the x-ray beam parallel to the magnetic field. Measurements were made by cooling the sample to 115±0.01 K (into the charge ordered phase) in the vapor from the liquid He reservoir at zero magnetic field. Measurements were conducted at fixed field with persistent currents.

Spectra were measured at the National Synchrotron Light Source (NSLS) at Brookhaven National Laboratory utilizing beam line X23B and a Si(111) monochromator and focusing torroidal mirror with an energy cutoff of 11 KeV were used. Spectra were taken in transmision mode using nitrogen filled ion chambers. A Mn foil reference was employed for energy calibration. The reduction of the XAFS data was performed using standard procedures [22]. Calibration was accomplished by defining the first inflection point in the spectrum of Mn metal as 6539 eV. The ionization threshold, $E_0$, was set at 6554.5 eV based on the zero field spectrum. The photoelectron momentum is defined as $k = \sqrt{(2m/\hbar^2)(E - E_0)}$. XAFS were extracted from the spectra as the difference between the normalized spectra and an adjustable spline function fit through the post-edge region, the parameters of which were adjusted to minimize low frequency residuals in the Fourier transform below R= 0.75 Å [23]. After comparison, two to four individual scans were averaged (see Fig. 1(a)).

Fits to the first shell Mn-O distribution were performed by Fourier transforming the XAFS data over the range $2.54 < k < 14.5$ Å$^{-1}$ to R-space [24]. The R-space fits (Fig. 2) to theoretical signals [25] based on the low temperature structure [26] of $La_{0.5}Ca_{0.5}MnO_3$ were performed over the range $0.75 < R < 1.95$ Å assuming a cumulant expansion [27] for the radial distribution (Fig. 1(b)). The simplest model which fit the data well consists of an amplitude reduction factor $S_0^2 = 0.7$ and variable parameters, R (average bond distance), $\sigma^2$ (the DW Factor ($<(R - <R>)^2>$ mean squared relative displacement) representing the width of the distribution) and C4 (kurtosis or flatness of the distribution $< (R - <R>)^4 > - 3\sigma^4$ ). The first shell coordination number was held fixed at 6. The total number of free parameters in the fit was 3, compared to the theoretical maximum number of independent parameters $2\Delta k\Delta r/p + 2 = 7$ [28]. Parameter errors were obtained by determining the stability of the fits with respect to changes in each parameter.

## III. Results and Discussion

Measurements were made at 115K, deep within the charge ordered ($T_N \sim 155$ K) region of this half doped system. Fourier transform spectra (qualitatively analogous to radial distributions with the caveat that the amplitude is scattering atom dependent and the peak position are shifted toward the origin by the phase of the scattering factor) of the field dependent XAFS data from 0 to 10 T are shown in Fig. 2 with the Mn-O, Mn-Ca/La and Mn-Mn/Mn-O-Mn peaks labeled. Note the enhancement of Mn-O peak with increasing magnetic field. This indicates that there is a reduction of the $MnO_6$ octahedral distortion with field.

A systematization of the rends with field can be obtained by looking at the higher order bond correlations, Note the strong enhancement of the Mn-Mn/Mn-O-Mn shoulder at high field (above 8.5 T). In Fig. 2(b) we expand the Mn-Mn/Mn-O-Mn peak to show that three distinct structural regions exist. For fields below ~7.5 T the low temperature charge ordered (AF) structure is maintained. Between ~8.5 and ~9.25 T a mixed-phase region exists. Above ~9.3 T, a ferromagnetic phase exists with structure similar to the high temperature paramagnetic phase (300 K, dotted line).

To quantify the changes in the local structure about the Mn site we have extracted the DW Factor ($\sigma^2 = <(R - <R>)^2>$) or variance of the Mn-O distribution as a function of the magnetic field. (Fig. 3). The results are compared with high field measurements of the magnetization by Xiao *et al* [5] (where the difference in the magnetization measured at 108 K and our 115K measurements are within the size of the solid dots). The moment at 10 T is 3.0 $\mu_B$/Mn site and changes by only 4 % (due to thermal disorder) between 10

T and 18 T [5]. The saturation moment of 3.5 $\mu_B$/Mn is not achieved even at these high fields. However at low temperature (below 77K) the saturation moment is achieved over the same field region indicating the moment loss is due to thermal disorder of the moments [5]. Since the moment loss is small, we make arguments concerning the saturation limit in $\sigma^2$ (H). Note that at high field the Jahn-Teller distortion is strongly suppressed. The high field values should be compared to the value $\sigma^2$ = 0.00254 0.00021 Å$^2$ found for CaMnO$_3$ (at 300K) with octahedral Mn-O distances of 1.895, 1.900 and 1.903 Å [29]. The data points are labeled near the region of rapid change in amplitude. As mentioned above, the distribution of Mn-O distances in zero field is non-Gaussian and is described by one additional parameter (C4 = < (R - <R>)$^4$ > - 3 $\sigma^4$) characterizing the flatness (due to multiple Mn-O bond lengths). In Fig. 4 we see that at high field C4 vanishes indicating the disappearance of the Jahn Teller distortion (resulting in a narrow Gaussian distribution of distances).

In Fig. 5 we show the average distance as a function of magnetic field, The average Mn-O bond distance approaches 1.96Å, the value measured in the optimally ferromagnetic region of the La$_{1-x}$Ca$_x$MnO3 phase diagram near x = 0.33 [30]. Again the increase in bond distance above 7.5 T signals the existence of a mixed phase region occurring before transformation to the ferromagnetic phase.

In Fig 6. we plot the DW Factor ($\sigma^2$) as a function of the magnetization. One can clearly see from this figure that in the charge ordered region that Jahn-Teller distortion decreases approximately linearly with the spin polarization. On crossing into the ferromagnetic metal (FMM) phase, the decrease in the distortion occurs at a rate 100 times higher than in the CO region.

The origin of the large decrease in resistivity observed in this system is now understood [5,8]. The JT distortion on the Mn site traps the $e_g$ electrons and reduces the mobile carrier density. Application of a magnetic field first polarizes (aligns) the Mn moments by destroying the CE AF phase leading to and exponential drop in resistivity. Application of a field beyond the first critical field then results in the collapse of the JT distortion concomitant with a change in the long range structure further reducing the resistivity. At high filed the $e_g$ electrons become un-trapped, with progressively lower binding and are able to mediate the ferromagnetic interactions between Mn sites.

We note that the reduction of the DW factor is of the order of 40% when comparing the 0 and 10 T values. This value matches well with the temperature dependent collapse of the DW factor in the ferromagnetic phase of $La_{1-x}Ca_xMnO_3$ (x~0.33) on going from the high temperature insulating and paramagnetic state to the low temperature ferromagnetic metallic state [18].

## IV. Summary

The local structure about the Mn site in the half doped system $La_{0.5}Ca_{0.5}MnO_3$ was measured in magnetic fields up 10 T to probe the melting of the charge ordered state. The Debye-Waller factor for the first shell Mn-O distribution is reduced by ~40 % between 0 and 10T. Examination of the Mn-O and Mn-Mn correlations reveal three distinct regions in the structure-field diagram. A broad region with weak field dependence (mainly antiferromatnetic phase below 7.5 T (CO Region)), a narrow-mixed phase region near ~ 8.5 T followed by a ferromagnetic phase region (FMM) with strong field-structure coupling. In terms of the magnetization, the variance of the Mn-O distribution is linear both in the CO and FMM regions with a 100-fold increase in slope in the FMM region. At high field the the Mn-O radial distribution becomes Gaussian and the Mn-Mn correlations are enhanced - consistent with the dominance of a ferromagnetic phase. The exponential change in resistivity in the first region is dominated by the reordering of the moments on the Mn sites from CE type antiferromagnetic to ferromagnetic order. This results suggest that spin scattering plays an important role in the magnetoresistance. The reduction in resistivity at high field is seen to be the result of the delocalization of JT trapped electrons as the magnetic field destabilizes the local distortions.

# Acknowledgments

Data acquisition was performed at Brookhaven National Laboratory's National Synchrotron Light Source (NSLS) which is funded by the U. S. Department of Energy. The magnet acquisition funded by NSF IMR Grant DMR-0083189 and operational costs were funded by NSF DMR-0209243. Sample preparation was supported by NSF DMR 9802513 (SWC). We are indebted to Dr. Steven Lamarra of NSLS for much technical assistance in the commissioning of the magnet.

# Captions

**Fig. 1.** (a) Extracted $La_{0.5}Ca_{0.5}MnO_3$ XAFS data for two consecutive scans at 115 K and zero field are displayed to show the level of the noise in the data. The first and second scan are displayed as thick and thin lines, respectively. (b) Typical fit to Mn-O shell over the range 0.75< R<1.95 Å with thin lines corresponding to the data and thick lines corresponding to the fit. The dashed and solid lines correspond to the magnitude and imaginary parts, respectively.

**Fig. 2.** (a) Fourier transform of the field dependent XAFS data (as in Fig. 1) over the range 2.54 < k < 14.05 Å$^{-1}$ with the Mn-O, Mn-Ca/La and Mn-Mn/Mn-O-Mn peaks labeled. (b) The expand Mn-Mn/Mn-O-Mn peak shows that three distinct structural regions exists. The 300 K spectrum is shown as the dotted line.

**Fig. 3.** The field dependent Debye-Waller factor ($\sigma^2 = <(R - <R>)^2>$, open circles) for Mn-O bond compared with the magnetization (solid dots taken from Xiao *et al.* [5]).

**Fig. 4.** The field dependent fourth cumulant (C4 = $<(R - <R>)^4> - 3\sigma^4$) for Mn-O bond. This quantity measures the flatness (kurtosis) of the radial distribution. As in the case of the Debye-Waller factor at high field C4 vanishes indicating the disappearance of the Jahn Teller distortion (resulting in a narrow Gaussian distribution of distances Fig. 4).

**Fig. 5.** The average Mn-O bond distance is seen to approaches 1.96 Å, the value measured in the optimally ferromagnetic region of the $La_{1-x}Ca_xMnO3$ phase diagram near $x = 0.33$.

**Fig. 6.** Debye-Waller factor plotted as a function of magnetization. The JT distortion varies linearly in both the metallic (FMM) and charge ordered (CO) regions. The rate of change of the JT distortion is 100 times faster after crossing into the ferromagnetic metallic region.

**Fig. 1.** Tyson *et al.*

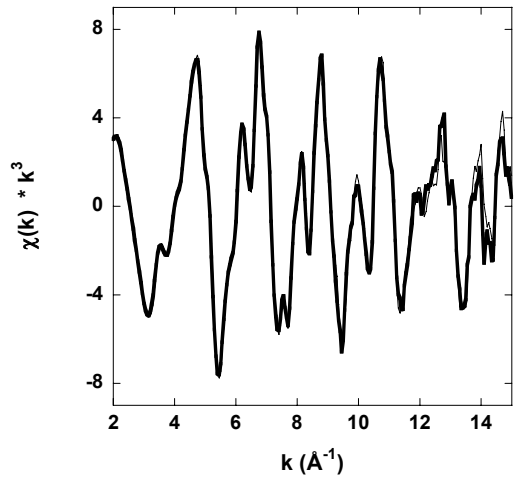

(a)

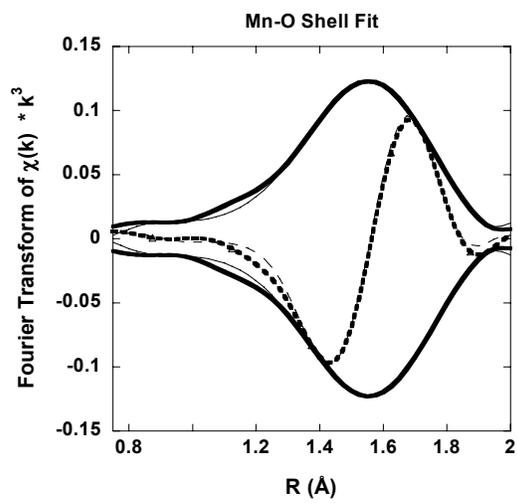

(b)

**Fig. 2.** Tyson *et al.*

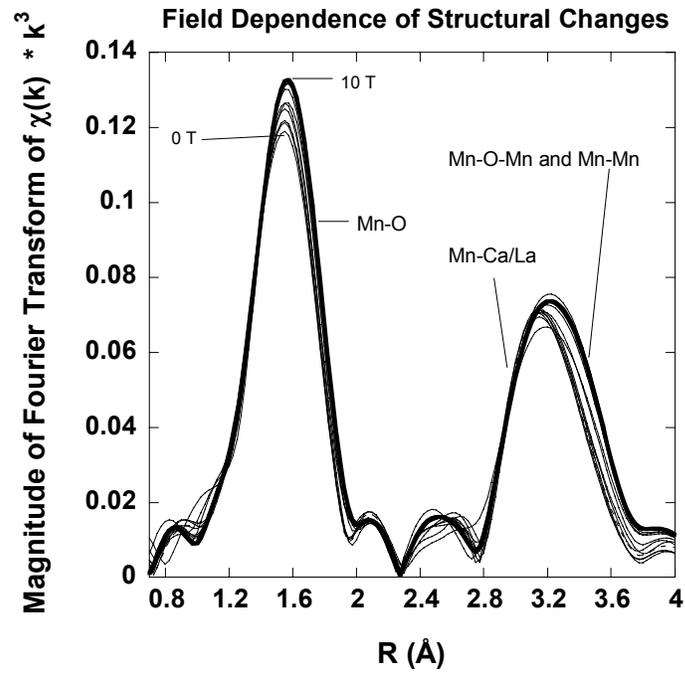

(a)

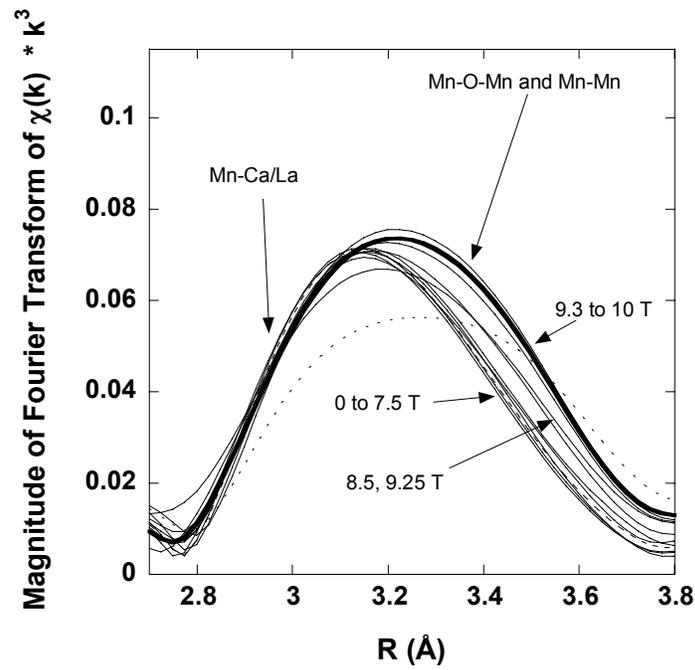

(b)

**Fig. 3.** Tyson *et al.*

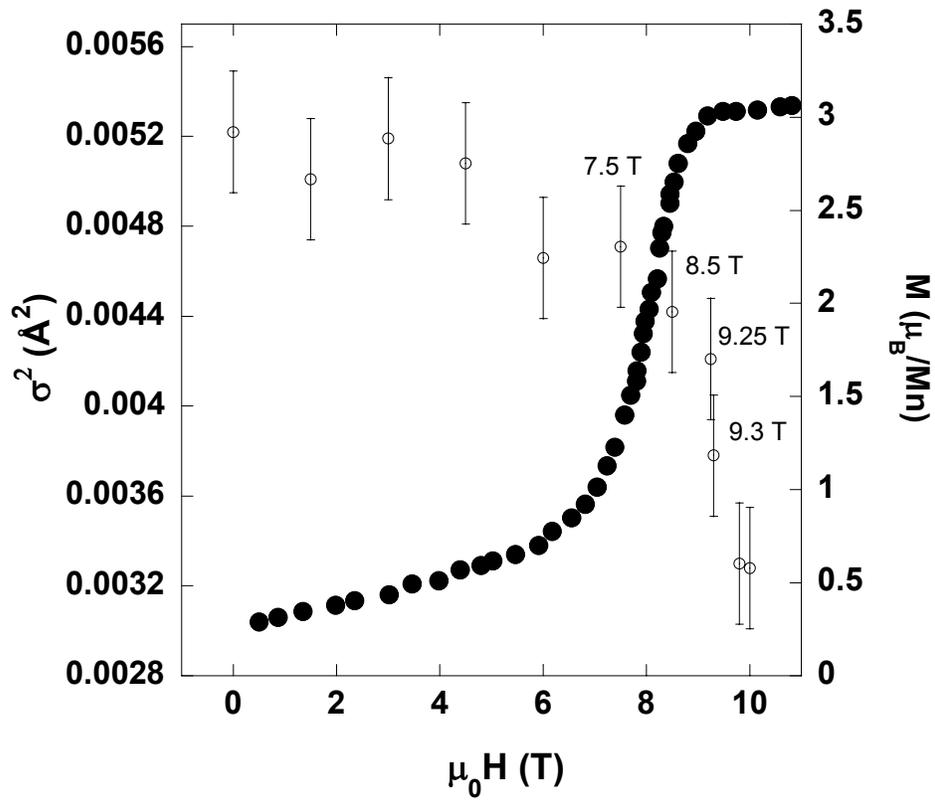

**Field Dependent Debye-Waller Factor and Magnetization**

**Fig. 4.** Tyson *et al.*

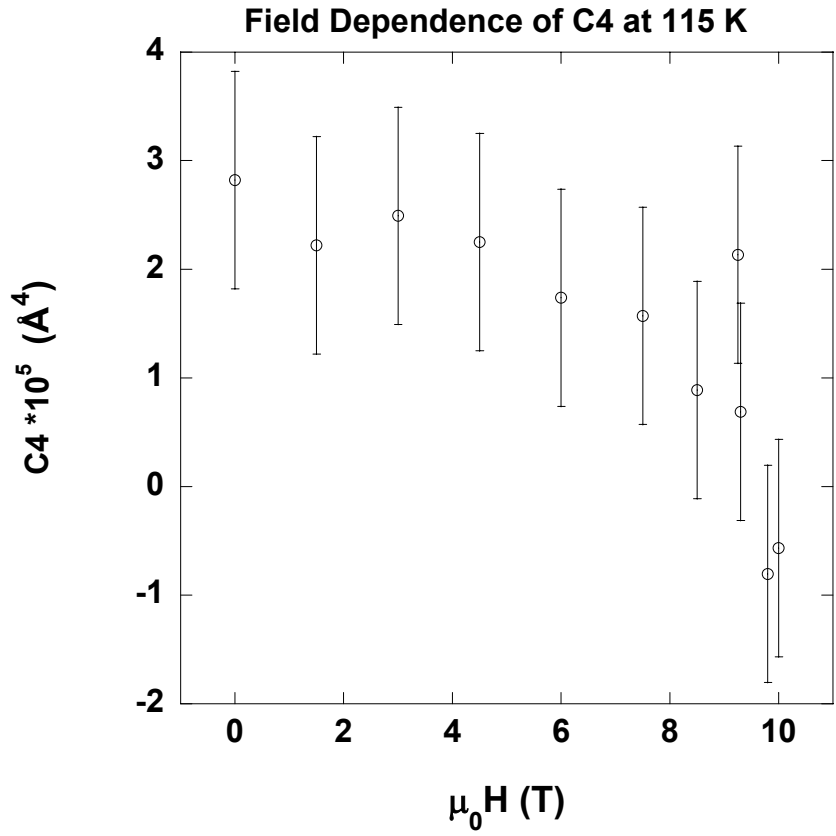

**Fig. 5.** Tyson *et al.*

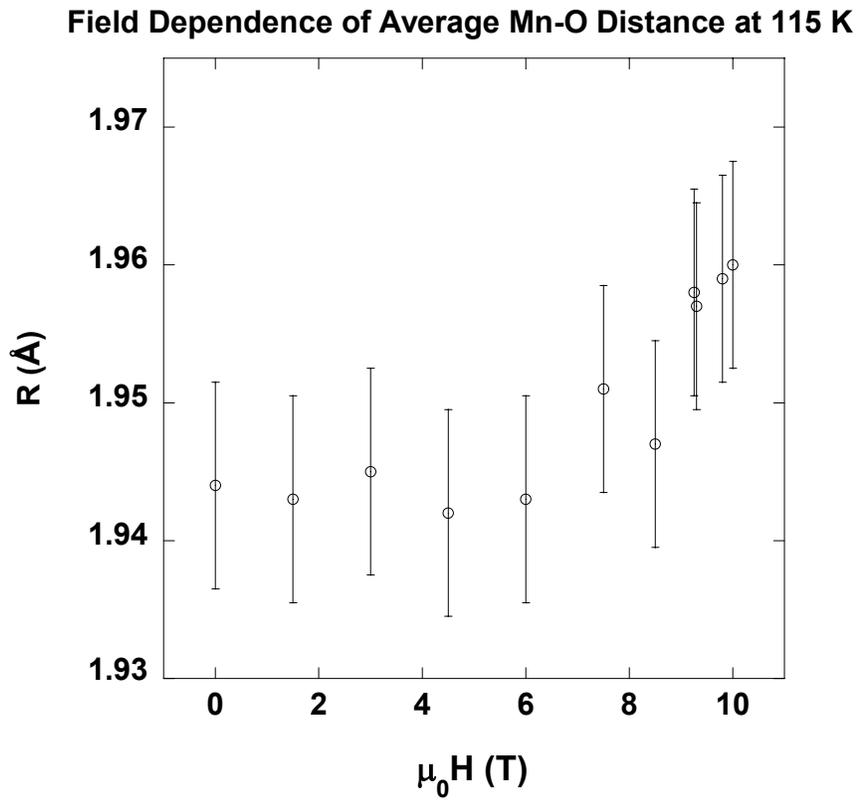

**Fig. 6.** Tyson *et al.*

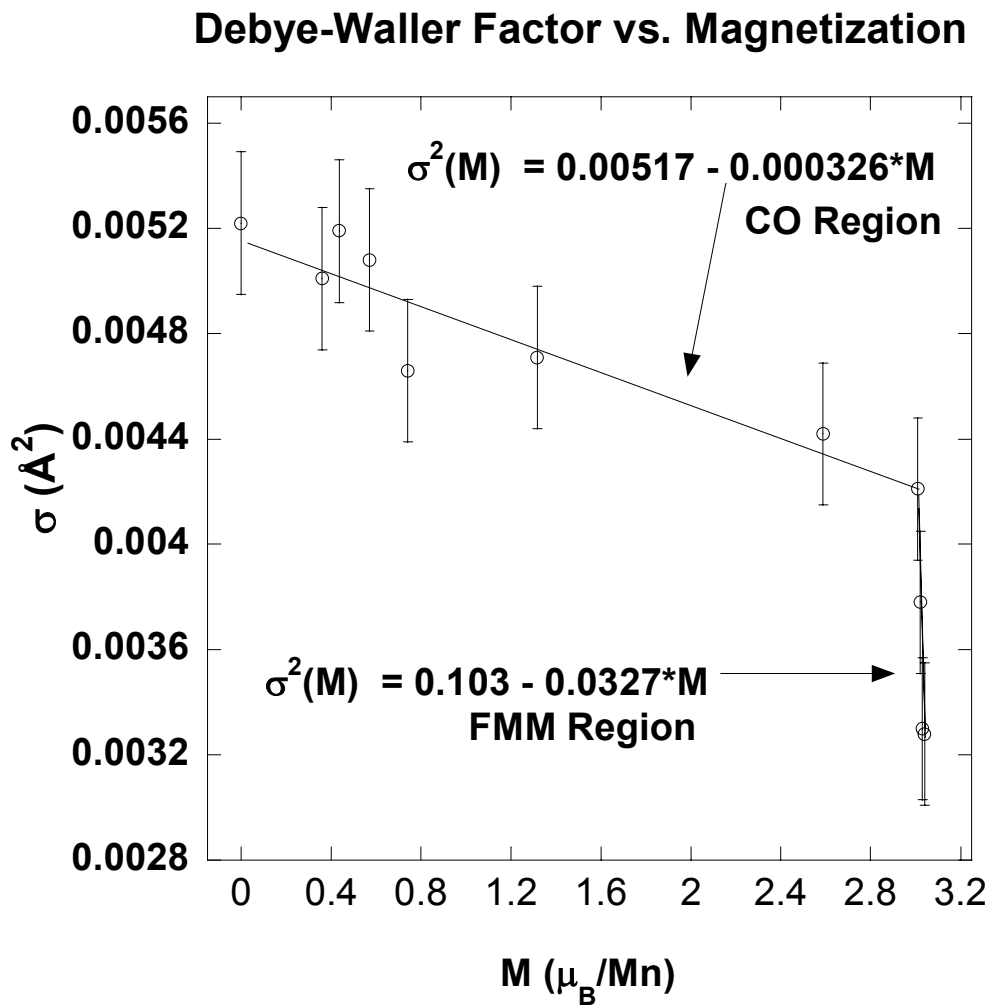

M ($\mu_B$/Mn)